\documentclass[aps,prl,preprint,groupedaddress,floatfix,floatfix]{revtex4-2}
\usepackage{graphicx}
\usepackage{bm}
\usepackage{hyperref}
\usepackage{amsmath}
\usepackage{amssymb}

\begin{document}

\title{Nonlinear spin-Wave Doppler effect for flexible tuning of magnonic frequencies}

\author{Jincheng Hou}
\thanks{These authors contributed equally to this work.}
%\email{jchou@hust.edu.cn}
\affiliation{College of Integrated Circuits and Optoelectronic Chips, Shenzhen Technology University, Shenzhen 518118, China}
\affiliation{School of Integrated Circuits, Huazhong University of Science and Technology, Wuhan 430074, China}

\author{Shaojie Hu}
\thanks{These authors contributed equally to this work.}
\email{hushaojie@sztu.edu.cn}
\affiliation{College of Integrated Circuits and Optoelectronic Chips, Shenzhen Technology University, Shenzhen 518118, China}

\author{Long You}
\email{lyou@hust.edu.cn}
\affiliation{School of Integrated Circuits, Huazhong University \\of Science and Technology, Wuhan 430074, China and \\Shenzhen Huazhong University of Science and Technology Research Institute, Shenzhen 518000, China and \\Key Laboratory of Information Storage System, Ministry \\of Education of China, Wuhan 430074, China}

\date{\today}

% Optional visible preprint statement (common for preprint but not required for submission)
%\noindent\textit{Note: The first two authors contributed equally and share co--first authorship.}

\begin{abstract}

We theoretically propose a nonlinear spin-wave Doppler effect, in which the time-dependent motion of a magnetic energy boundary acts as an active frequency modulator, directly converting boundary-induced phase dynamics into instantaneous spectral synthesis for propagating spin-wave modes. 
In contrast to the conventional linear Doppler effect governed by constant relative velocity, this mechanism enables dynamic phase-to-frequency transduction, generating high-order harmonics, magnonic frequency combs, and coherent chirped sidebands, without requiring nonlinear magnon–magnon coupling or multi-magnon scattering. 
Micromagnetic simulations on voltage-controlled anisotropy boundaries in ferroelectric/ferromagnetic (FE/FM) heterostructures demonstrate that the comb spacing and spectral topology are determined solely by boundary kinematics, confirming direct Doppler phase coupling between boundary motion and spin-wave propagation. 
These results establish moving magnetic-energy boundaries as a new class of on-chip spectral synthesizers and define a coherent and energy-efficient framework for flexible tuning of magnonic frequencies, fundamentally distinct from traditional passive scattering or nonlinear multi-magnon mechanisms.

\end{abstract}

\maketitle

\newpage
\section{Introduction}

Spin waves (SWs), collective excitations of precessing magnetic moments arising from exchange and dipolar interactions, have emerged as promising information carriers for next-generation spintronic and magnonic technologies\cite{SW_rev2_HanXF,SW_rev3_RPTU,sw_rev4,SW_rev_coh}. Their key advantages—charge-free propagation\cite{SW_JouleLess_1,SW_JouleLess_2,SW_JouleLess_3}, ultralow power consumption\cite{SW_JouleLess_4,sw_amp_1,sw_amp_2,hou2025excitation}, nanoscale wavelength\cite{SW_shortWlen_1,SW_shortWlen_2,SW_Harmony_8}, and wide frequency bandwidth\cite{sw_bandwh_1,sw_bandwh_2,sw_bandwh_3,SW_FrequencyComb_6}—enable applications ranging from microwave and terahertz signal processing\cite{Yu2025,sw_amp_1,sw_amp_2,sw_bandwh_1,sw_bandwh_2,SW_FrequencyComb_6} to in-memory logic\cite{sw_rev4,SW_logic_1,SW_logic_2,SW_logic_3} and quantum information devices\cite{SW_qbit,SW_qbit_1}. These opportunities have stimulated intense efforts to realize robust and reconfigurable frequency control of SWs, a central requirement for magnon-based circuitry.
Conventional approaches to SW frequency tuning—static magnetic fields\cite{Yu2025,SW_logic_2,SW_logic_3,sw_fieldtune}, materials engineering\cite{SW_Harmony_10,SW_Harmony}, spin-texture manipulation\cite{sw_bandwh_2,SW_FrequencyComb_1,SW_FrequencyComb_2,SW_FrequencyComb_3,SW_FrequencyComb_4,SW_Harmony_1,SW_Harmony_2,SW_Harmony_5,SW_Harmony_6,SW_Harmony_7}, and current-driven modulation\cite{SW_Harmony_6,sno_1}—typically offer limited dynamical tunability, finite bandwidth, or high energy overhead. The spin-wave Doppler effect\cite{SW_Doppler_STT_0}, arising from the relative motion between SWs and magnetic structures, provides an appealing alternative route. Previous studies have revealed Doppler shifts induced by current-driven spin-transfer torque\cite{SW_Doppler_STT_0,SW_Doppler_STT_1,SW_Doppler_STT_2}, magnon drag\cite{SW_Doppler_1}, motion of domain walls\cite{SW_Doppler_DW_1,SW_Doppler_DW_2}, movable magnetic boundaries\cite{SW_Doppler_2,zhong2025control}, and source–detector motion in magnonic waveguides\cite{SW_Doppler_5,SW_Doppler_6}. However, these mechanisms are all governed by uniform or linear motion and therefore produce only simple red– or blue-shifts in SW frequency. Such linear Doppler effects are insufficient for continuous, broadband, or strongly nonlinear modulation—capabilities essential for advanced magnonic frequency processors.

In this Letter, we introduce the nonlinear spin-wave Doppler effect, a previously unexplored mechanism in which a time-dependent, non-uniformly moving magnetic energy boundary acts as an active frequency modulator for SWs. Unlike linear-motion-induced Doppler shifts, nonlinear boundary motion generates qualitatively new phenomena, including magnonic frequency combs, higher-order harmonics, and pronounced frequency chirps. These effects cannot be attributed to multi-magnon coupling or previously reported nonlinear magnonic processes\cite{SW_FrequencyComb_1,SW_FrequencyComb_2,SW_FrequencyComb_3,SW_FrequencyComb_4,SW_FrequencyComb_5,SW_FrequencyComb_6,SW_Harmony,SW_Harmony_1,SW_Harmony_2,SW_Harmony_3,SW_Harmony_4,SW_Harmony_5,SW_Harmony_6,SW_Harmony_7,SW_Harmony_8,SW_Harmony_10}.
Using micromagnetic simulations, we demonstrate that periodically accelerated magnetic anisotropy boundaries generate a broad set of SW harmonics and a well-defined frequency comb—a distinctive signature of unconventional Doppler modulation. Our results reveal a conceptually new regime of magnon–boundary interaction and establish nonlinear Doppler physics as a powerful platform for compact, tunable, and ultralow-power magnonic frequency modulation.

\section{Theory}

To reveal the mechanism underlying the nonlinear Doppler effect induced by a time-dependent magnetic energy boundary (MEB), we consider a spin wave interacting with a boundary whose velocity contains uniform, accelerating, and oscillatory components,
\begin{equation}
\bm{v}(t)=\bm{v}_{0}+\bm{a}(t-t_{0})+\bm{V}\cos(\Omega t).
\label{eq:velocity}
\end{equation}

Under the quasistatic condition ($|\bm{v}|\!\ll\!v_{g}$ and $\omega_{0}\tau_{c}\!\gg\!1$, where $v_{g}$ is the group velocity of the incident spin wave and $\tau_c$ is the effective interaction time determined by the overlap between the moving boundary and the spin-wave packet), the instantaneous frequency of the reflected or transmitted wave in the laboratory frame follows from a first-order Doppler expansion,
\begin{equation}
\omega_{r,t}(t)
=\omega_{0}-\Delta\bm{k}(\omega_{r,t})\!\cdot\!\bm{v}(t),
\label{eq:Doppler_general}
\end{equation}
where $\omega_{0}$ is the incident frequency and $\Delta \bm{k}(\omega)=\bm{k}_{0}(\omega)-\bm{k}'_{r,t}(\omega)$ denotes the wave-vector difference across the moving boundary. $\Delta\bm{k}_{0}=\Delta\bm{k}(\omega_{0})$ is the static wave-vector mismatch across the boundary for $\omega=\omega_{0}$.  
The dynamical magnetization takes the form
\begin{equation}
\psi_{r,t}(x,t)
=A_{r,t}(t)\exp\!\left[i\big(\bm{k}'_{r,t}\!\cdot\!\bm{x}-\Phi_{r,t}(t)\big)\right],
\label{eq:wavefunction}
\end{equation}
where $A_{r,t}(t)$ is the slowly varying amplitude envelope, and the instantaneous phase is given by 
$\Phi_{r,t}(t)=\int^{t}\omega_{r,t}(t')\,dt'+\phi_{r,t}^{(0)}$, with $\phi_{r,t}^{(0)}$ being the initial phase constant at $t=t_{0}$. It should be noted that although the time-dependent envelope $A_{r,t}(t)$ introduces only a weak amplitude--frequency modulation through the velocity-dependent reflection or overlap factors, the accumulated phase $\Phi_{r,t}(t)$ overwhelmingly dominates the resulting frequency modulation because of the nonlinear Doppler effect.

Substituting Eq.~(\ref{eq:velocity}) into Eq.~(\ref{eq:Doppler_general})  and integrating over time yields the total accumulated phase
\begin{equation}
\Phi_{r,t}(t)-\omega_{0}t
=-\Delta\bm{k}_{0}\!\cdot\!\!\left[
\bm{v}_{0}(t-t_{0})
+\tfrac{1}{2}\bm{a}(t-t_{0})^{2}
+\frac{\bm{V}}{\Omega}\sin(\Omega t)
\right]+\phi_{r,t}^{(0)}.
\label{eq:Ins.phase}
\end{equation}
Equation~(\ref{eq:Ins.phase}) shows that acceleration generates a quadratic phase term, whereas oscillatory boundary motion produces sinusoidal phase modulation.  
These two contributions give rise to distinct nonlinear Doppler signatures.

\begin{figure}[htb]
	\centering
	\includegraphics[width=6in]{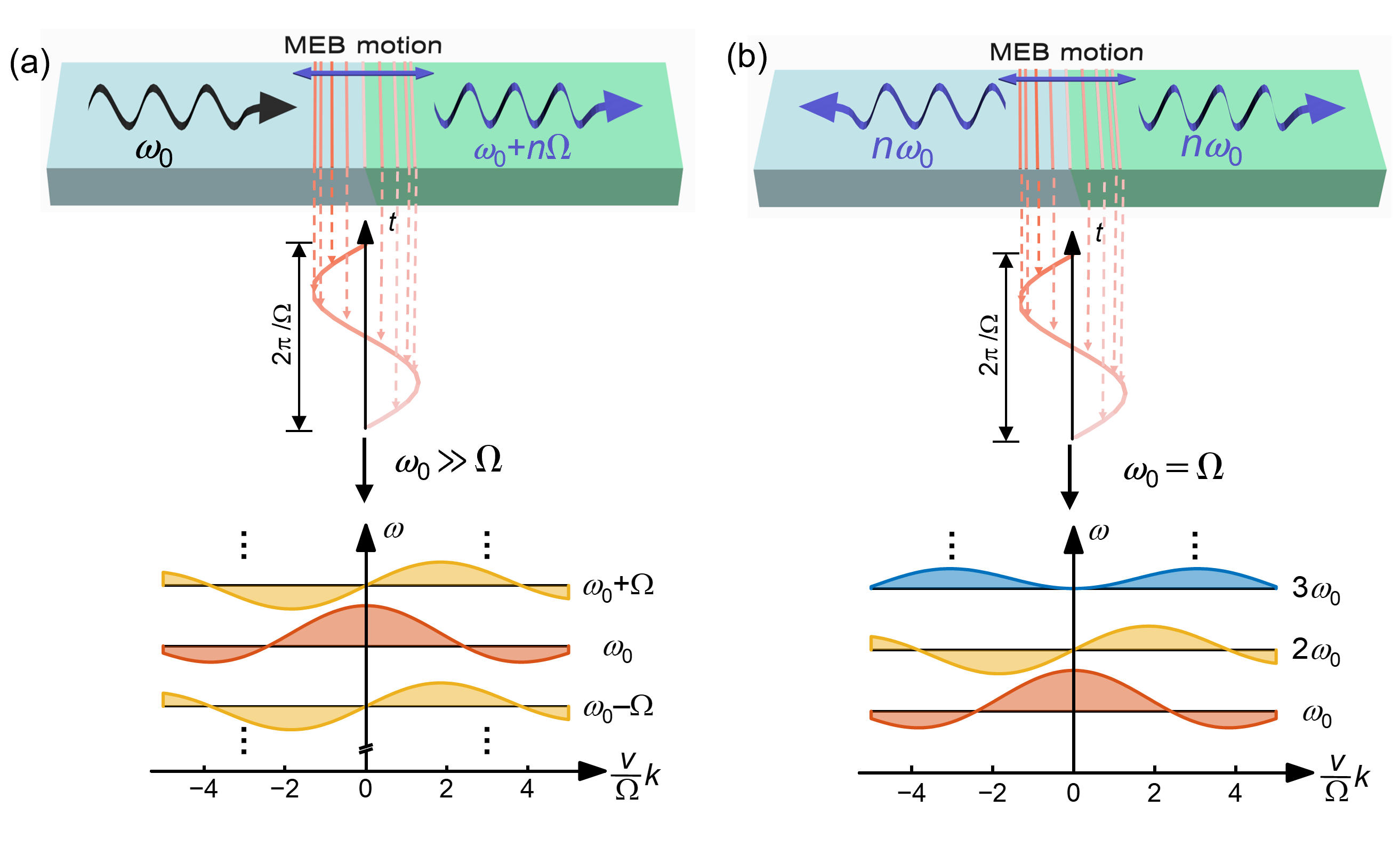}
    \caption{Illustration of nonlinear spin-wave Doppler effect.  (a) A spin wave of frequency $\omega_0$ interacting with a periodically oscillating MEB at $\Omega$ experiences a nonlinear Doppler shift, producing transmitted components at $\omega = \omega_c + n\Omega$ ($n \in \mathbb{Z}$) and forming a magnonic frequency comb.
    (b) A periodically oscillating MEB, which acts as a spin-wave emitter with $\omega_{0}=\Omega$, experiences nonlinear Doppler shifts, producing higher harmonics components at $\omega = n\omega_0$ ($n \in \mathbb{Z}$).
    }
	\label{Fig1}
\end{figure}

To further clarify the effect of periodic boundary motion, we examine how frequency combs and higher harmonics emerge.
When the oscillatory component dominates ($|\bm{a}|\!\ll\!|\bm{V}|$), the spin-wave solution reduces to
\[
\psi_{r,t}(x,t)\simeq 
A_{r,t}(t)\,e^{i(\bm{k}'_{r,t}\!\cdot\!\bm{x}-\omega_{c}t)}
e^{-i\beta\sin(\Omega t)},
\]
where the carrier frequency is
\[
\omega_{c}=\omega_{0}-\Delta\bm{k}_{0}\!\cdot\!\bm{v}_{0},
\]
and the modulation index is
\begin{equation}
\beta=\frac{\Delta\bm{k}_{0}\!\cdot\!\bm{V}}{\Omega}.
\end{equation}
The Jacobi–Anger expansion gives
\begin{equation}
\psi_{r,t}(x,t)=
A_{r,t}(t)\,e^{i(\bm{k}'_{r,t}\!\cdot\!\bm{x}-\omega_{c}t)}
\sum_{n=-\infty}^{\infty}J_{n}(\beta)e^{-in\Omega t},
\label{eq:FM_expansion}
\end{equation}
producing sidebands at $\omega=\omega_{c}+n\Omega$ ($n\in\mathbb{Z}$), as shown Fig.\ref{Fig1}(a).  
Thus, oscillatory MEB motion generates a magnonic frequency comb with intensities
\begin{equation}
I_{n}^{(r,t)}\propto |A_{r,t}|^{2}J_{n}^{2}(\beta).
\label{eq:Intensity}
\end{equation}
When $\omega_{0}=\Omega$, the oscillating boundary acts as an emitter, yielding higher harmonics at $\omega=n\Omega$, as shown Fig.\ref{Fig1}(b).

In contrast to the periodic-motion case, accelerated boundary motion yields a different type of spectral modulation.
If the acceleration term dominates ($|\bm{V}|=0$), the instantaneous frequency becomes
$\omega_{r,t}(t)=\omega_{c}-\Delta\bm{k}_{0}\!\cdot\!\bm{a}(t-t_{0})$,
leading to a linear frequency sweep.  
The chirp bandwidth over an observation time $T$ is
\begin{equation}
\Delta f_{\mathrm{chirp}}
\simeq \frac{|\Delta\bm{k}_{0}\!\cdot\!\bm{a}|\,T}{2\pi}.
\label{eq:chirpBW}
\end{equation}

In theory, nonlinear boundary motion generates a time-dependent Doppler phase shift that governs the observed spectral features.  
Periodic motion produces a Bessel-type frequency comb with spacing $\Omega$, whereas acceleration induces a broadband chirp proportional to $|\Delta\bm{k}_{0}\!\cdot\!\bm{a}|$.  
Thus, oscillation determines the frequency combs and higher harmonics, while acceleration controls the chirped frequency shifts, establishing the nonlinear Doppler effect as a versatile mechanism for magnon frequency modulation.

\newpage
\section{Micromagnetic Simulations}

\begin{figure}[htb]
	\centering
	\includegraphics[width=6in]{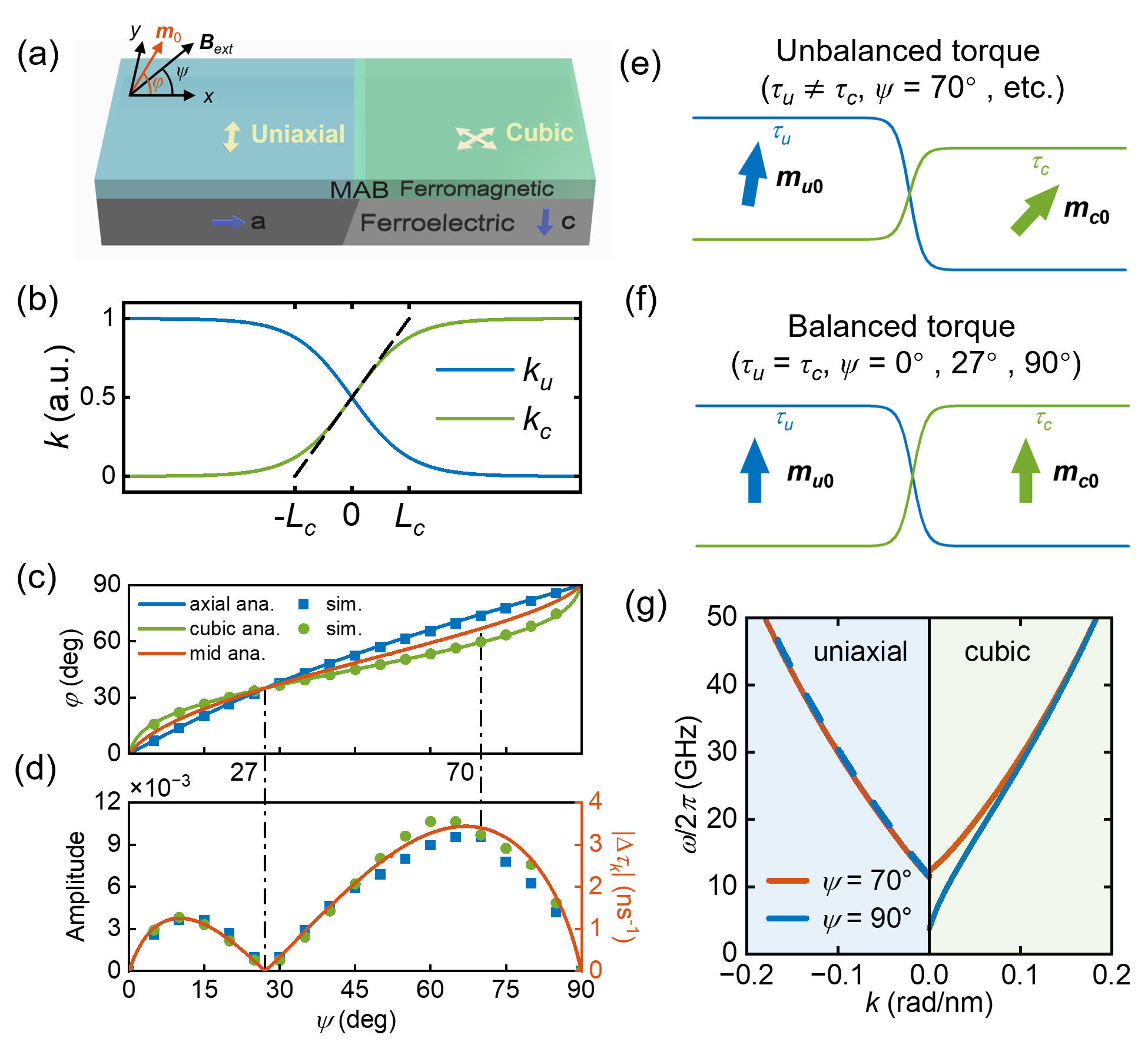}
    \caption{
    (a) Schematic illustration of the magnetic anisotropy boundary (MAB) formed by ferroelectric (FE) domains in a ferroelectric/ferromagnetic (FE/FM) heterostructure. In-plane \textit{a}-type FE domains induce uniaxial magnetic anisotropy ($K_{u0}$), while out-of-plane \textit{c}-type FE domains induce cubic magnetic anisotropy ($K_{c0}$). 
    (b) The spatial anisotropy distribution across the MAB follows a hyperbolic tangent profile. 
    (c) Static equilibrium magnetization angle $\varphi$ as a function of the external field angle $\psi$ between the two magnetic regions under $B_0 = 0.06~\mathrm{T}$. Squares and dots denote simulation data, and solid lines represent theoretical results. 
    (d) Anisotropy torque difference $|\Delta\tau_k|$ as a function of the external field angle $\psi$. 
    (e) Schematic of unbalanced anisotropy torque at $\psi = 70^{\circ}$. 
    (f) Schematic of balanced anisotropy torque at $\psi = 90^{\circ}$. 
    (g) Spin-wave dispersion relations under $\psi = 70^{\circ}$ and $\psi = 90^{\circ}$, calculated using Eqs.~(\ref{dpKu}) and (\ref{dpKc}) for uniaxial and cubic regions under an external magnetic field of $B_0 = 0.06~\mathrm{T}$.
    }
	\label{Fig2}
\end{figure}

To test the theoretical predictions, we performed micromagnetic simulations of a voltage-driven magnetic anisotropy boundary (MAB) in a BaTiO$_3$/Fe heterostructure\cite{FE_DomainMotion,FE_Domain_1} using \textsc{MuMax3}\cite{mumax}.  
The MAB position varies dynamically under an RF electric field applied to the ferroelectric layer, as illustrated in Fig.~\ref{Fig2}(a).  
Material parameters for Fe were  
$M_s=1.7\times10^6~\mathrm{A/m}$,  
$A_\mathrm{ex}=2.1\times10^{-11}~\mathrm{J/m}$,  
$\gamma=1.76\times10^{11}~\mathrm{rad/(T\cdot s)}$,  
$\alpha=0.01$,  
$K_{u0}=1.5\times10^4~\mathrm{J/m^3}$,  
$K_{c0}=4.4\times10^4~\mathrm{J/m^3}$.  
The uniaxial–cubic anisotropy transition follows the standard hyperbolic profile shown in Fig.~\ref{Fig2}(b):
\begin{equation}
K_u(x)=\tfrac{K_{u0}}{2}[1-\tanh((x-x_\mathrm{MAB})/L_c)],\quad
K_c(x)=\tfrac{K_{c0}}{2}[1+\tanh((x-x_\mathrm{MAB})/L_c)],
\end{equation}
where $L_c$ is the ferroelectric correlation length\cite{FE_Domain_2,FE_Domain_3}.

%%%%%%%%%%%%%%%%%%%%%%%%%%%%%%%%%%%%%%%%%%%%%%%%%%%%%%%%%%%%%%%%%%%%%%%%%%%%%
% Unbalanced torque: nonlinear emission
%%%%%%%%%%%%%%%%%%%%%%%%%%%%%%%%%%%%%%%%%%%%%%%%%%%%%%%%%%%%%%%%%%%%%%%%%%%%%

%\vspace{4pt}

We first evaluated the equilibrium magnetization angle $\varphi$ and the anisotropy-torque difference $|\Delta\tau_k|=\gamma\,\mathbf{m}\times(\mathbf{B}_u-\mathbf{B}_c)=\hat{z}\gamma\frac{\mathrm{sin}2\varphi_0}{M_S}\left(K_{u0}-K_{c0}\mathrm{cos}2\varphi_0\right)$  under an external field $B_0=0.06~\mathrm{T}$, shown in Fig.~\ref{Fig2}(c,d).  
The torque exhibits a pronounced maximum near $\psi=70^\circ$, corresponding to an \emph{unbalanced anisotropy torque}, where the MAB could serve as an efficient spin-wave emitter, shown in Fig. \ref{Fig2}(e). At $\psi=0^{\circ}$, $27^{\circ}$, and $90^{\circ}$, $|\Delta\tau_k|\rightarrow 0$, representing a \emph{balanced torque}, where it could act as a frequency modulator without spin-wave excitation, shown in Fig. \ref{Fig2}(f). 
And, the dispersion relations under $\psi=70^\circ$ and $\psi=90^\circ$ are also simulated both in uniaxial and cubic regions, as shown in Fig. \ref{Fig2}(g). The theory's calculations matched the simulation results well using the following equations\cite{SW_DispersionTheory,SW_DispersionTheory_2}.
\begin{equation}
    \omega_u=g_u\left(k\right)=\gamma\sqrt{\left(B_\eta-\frac{2K_{u0}}{M_S}\mathrm{cos}\left(2\varphi\right)\right)\left(B_\xi+\frac{2K_{u0}}{M_S}\right)+F_g} 
    %\tag{10a} 
    \label{dpKu}        
\end{equation}
\begin{equation}
    \omega_c=g_c\left(k\right)=\gamma\sqrt{\left(B_\eta-\frac{2K_{c0}}{M_S}\mathrm{cos}\left(4\varphi\right)\right)\left(B_\xi+\frac{K_{c0}}{M_S}\left(2+\mathrm{cos}\left(2\varphi\right)\right)\right)+F_g} 
    %\tag{10b} 
    \label{dpKc}
\end{equation}
where $B_\eta=B_0\mathrm{cos}\left(\varphi-\psi\right)+\frac{2A_{ex}}{M_S}k^2$, $B_\xi=B_0\frac{\mathrm{sin}\psi}{\mathrm{sin}\varphi}+\frac{2A_{ex}}{M_S}k^2+\mu_0M_S\left(1-G_k\mathrm{cos}^2\varphi\right)$, $F_g=G_k\left(1-G_k\right)\left(\mu_0M_S\mathrm{sin}\varphi\right)^2$, $G_k=1-\frac{1-e^{-\left|k\right|d_t}}{\left|k\right|d_t}$, $d_t$ is the thickness of the ferromagnetic layer.

\begin{figure}[htb]
	\centering
	\includegraphics[width=6in]{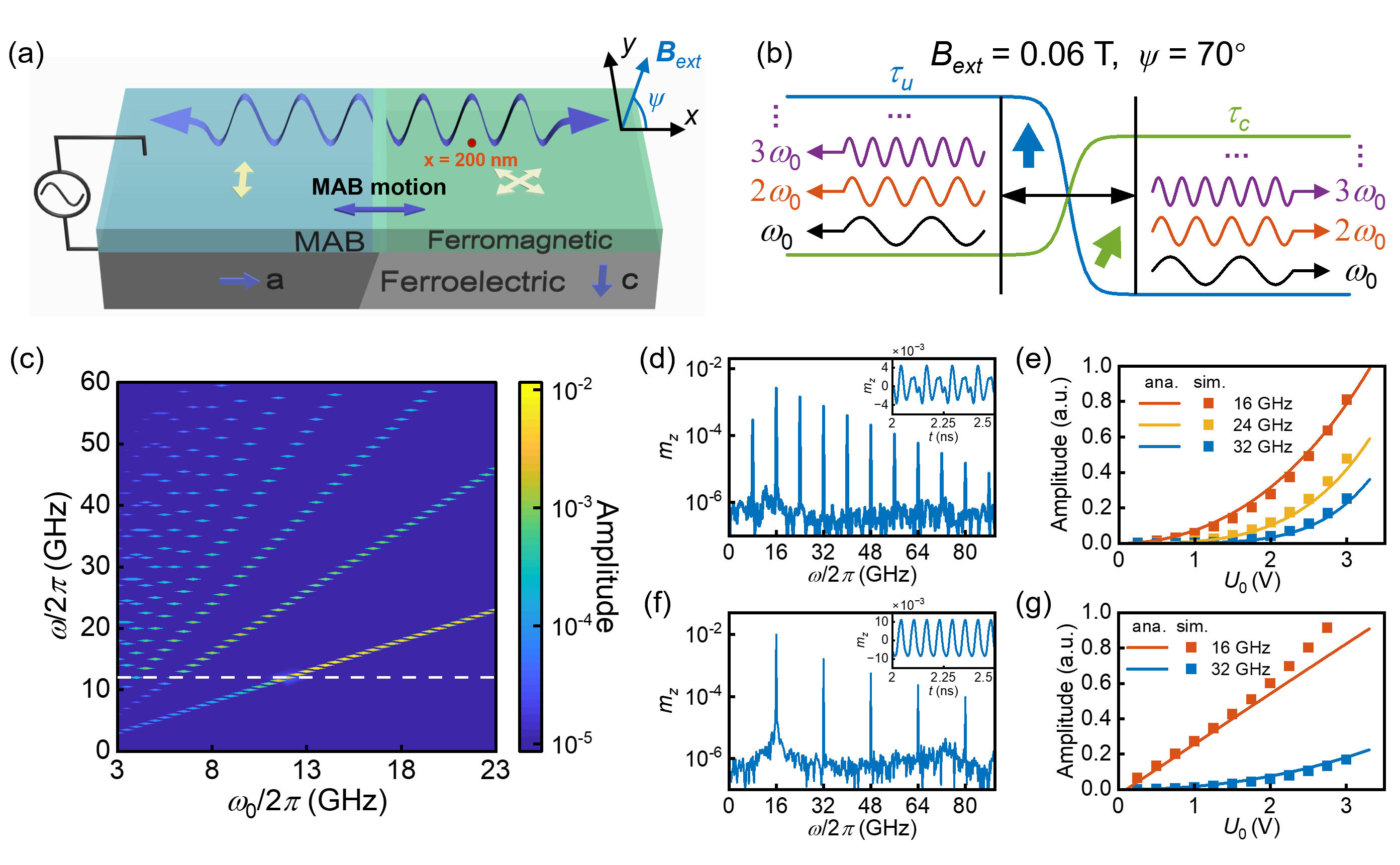}
    \caption{
    (a) Schematic illustration of a voltage-controlled MAB acting as a spin-wave emitter in an FE/FM heterostructure. 
    (b) Schematic illustration of harmonic spin-wave generation induced by the voltage-controlled unbalanced anisotropy torque. 
    (c) Contour plot of excited spin-wave spectra with various excitation frequencies from $3$ to $23~\mathrm{GHz}$. 
    (d,f) Fourier spectra and corresponding time-dependent magnetization (insets) at excitation frequencies of $8~\mathrm{GHz}$ and $16~\mathrm{GHz}$, respectively. 
    (e,g) Intensities of high-order harmonics under excitation frequencies of $8~\mathrm{GHz}$ and $16~\mathrm{GHz}$, respectively. Squares denote simulation data, and solid lines represent theoretical results. 
    }
	\label{Fig3}
\end{figure}

To reveal the effect of unbalanced anisotropy torque, we consider spin-wave emission and the resulting higher harmonic components.
The spin-wave excitation induced by the nonlinear spin-wave Doppler effect, utilizing an unbalanced anisotropy torque under an external magnetic field of $B_0 = 0.06~\mathrm{T}$ and an angle $\psi = 70^{\circ}$, as shown in Fig.~\ref{Fig3}(a).
Under this unbalanced torque, an RF voltage of $U_0=3~\mathrm{V}$ drives the MAB to a maximum velocity of $v_0\approx 966~\mathrm{m/s}$\cite{FE_Domain_1,FE_speed_1}.  
When the excitation frequency is set to $\omega_0 / 2\pi$, the simulated spatial distribution of the spin waves shows symmetric emission from the oscillating MAB, with nearly identical amplitudes but a phase difference on either side, as depicted in Fig.~\ref{Fig3}(b). The time-dependent magnetization, related to the phase of the spin waves, was measured at $x = 200~\mathrm{nm}$. This magnetization exhibits periodic oscillations, with its Fourier spectrum containing a series of equidistant harmonics, spaced by $\omega_0 / 2\pi$.
The complete contour plot of the excitation frequency response spectra, ranging from $3$ to $23~\mathrm{GHz}$, is shown in Fig.~\ref{Fig3}(c). Notably, when $\omega_0$ is lower than the ferromagnetic resonance frequency $\omega_\mathrm{FMR}$, higher-order harmonics continue to propagate, while the fundamental component below $\omega_\mathrm{FMR}$ decays rapidly with distance.
Figures~\ref{Fig3}(d) and (f) show the Fourier spectra and time-dependent magnetization (insets) at excitation frequencies of $8~\mathrm{GHz}$ and $16~\mathrm{GHz}$. Since $\omega_0 / 2\pi = 8~\mathrm{GHz}$ is less than $\omega_\mathrm{FMR} / 2\pi = 12~\mathrm{GHz}$, the strongest excitation occurs at the second harmonic frequency ($16~\mathrm{GHz}$). When $\omega_0 / 2\pi$ is increased to $16~\mathrm{GHz}$, the harmonic sequence begins from the first order, with a spacing of $16~\mathrm{GHz}$. These results demonstrate that the harmonic spacing is determined solely by the MAB oscillation frequency $\omega_0$, confirming that nonlinear boundary motion directly governs the spin-wave spectral structure.
We also simulated the intensities of high-order harmonics under the excitation frequency of $8~\mathrm{GHz}$ and $16~\mathrm{GHz}$ , as shown in Figs.~\ref{Fig3}(e) and (g). The simulated dependence of harmonic intensity on $U_0$ agrees well with the theoretical harmonic amplitudes $I_n$ calculated using Eq.~\eqref{eq:Intensity}.

\begin{figure}[htb]
	\centering
	\includegraphics[width=6in]{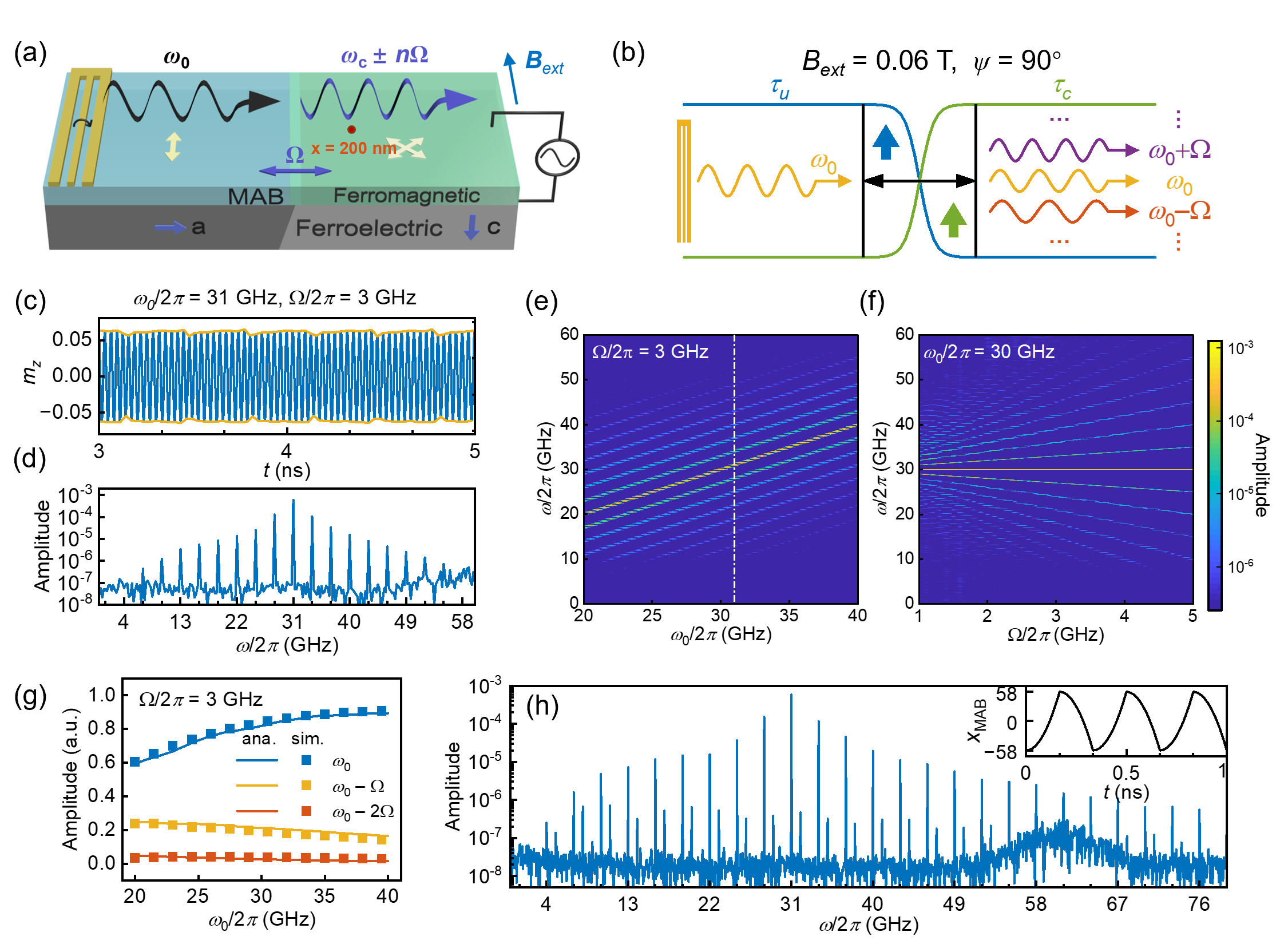}
    \caption{  
    (a) Schematic illustration of a voltage-controlled MAB acting as a spin-wave modulator in a FE/FM heterostructure. 
The incident spin wave has a frequency of $\omega_0 = 31~\mathrm{GHz}$, while an applied RF voltage of $\Omega/2\pi = 3~\mathrm{GHz}$ drives a periodic motion of the MAB at the same frequency. 
(b) Conceptual schematic of frequency-comb generation via the spin-wave nonlinear Doppler effect under a voltage-controlled balanced anisotropy torque. 
(c,d) Time-dependent oscillation of $m_z$ and its corresponding Fourier transform, measured at $x = 200~\mathrm{nm}$. 
(e) Contour plot of transmitted spin-wave spectra for various incident frequencies at a fixed modulation frequency $\Omega/2\pi = 3~\mathrm{GHz}$. 
(f) Contour plot of transmitted spin-wave spectra for various modulation frequencies $\Omega/2\pi$ at a fixed incident frequency $\omega_0/2\pi = 30~\mathrm{GHz}$. 
(g) Amplitudes of the main mode and two principal side peaks as a function of $\omega_0$. Squares denote simulation results, and solid lines represent theoretical predictions. 
(h) Transmitted spin-wave spectrum under a periodically accelerated motion of the MAB (inset), with the incident spin-wave frequency $\omega_0/2\pi = 31~\mathrm{GHz}$.
    }
	\label{Fig4}
\end{figure}

In contrast to the unbalanced-torque regime, the balanced anisotropy torque enables spin-wave modulation with frequency-comb features. Accordingly, we simulated the spin-wave modulation with spin-wave nonlinear Doppler effect by using balanced anisotropy torque under $B_0=0.06~\mathrm{T}$ and $\psi=90^{\circ}$, as illustrated in Figure~\ref{Fig4}(a,b).  
When the torque is balanced, the MAB does not generate spin waves but instead acts as a pure phase modulator. 
The time-dependent magnetization (related phase of transmitted spin-wave) recorded at $x=200~\mathrm{nm}$ exhibits a tiny amplitude envelope with a period of $1/\omega_0$ in Fig.~\ref{Fig4}(c)]. % due to the nonlinear dispersion of the medium. 正常调频或者调相是不会产生包络的，这里的包络是非线性色散的结果
While, its Fourier spectrum shows a well-defined frequency comb centered at $31~\mathrm{GHz}$ with spacing of $3~\mathrm{GHz}$ in Fig.~\ref{Fig4}(d). This demonstrates that periodic boundary motion effectively modulates the spin-wave frequency without generating additional excitation sources.

One of the most significant advantages of this mechanism lies in the independent tunability of $\omega_0$ and $\Omega$. As shown in Fig.~\ref{Fig4}(e), varying the modulation frequency $\Omega$ at a fixed input $\omega_0=30~\mathrm{GHz}$ continuously adjusts the sideband spacing, enabling wideband frequency control. Conversely, fixing $\Omega=3~\mathrm{GHz}$ and sweeping $\omega_0$  demonstrates uniform modulation behavior across a broad frequency range in Fig.~\ref{Fig4}(f), confirming the robustness of the spin-wave nonlinear Doppler effect.
The theoretical amplitudes $I_n = A(\omega_0)J_n[(k'_c - k_u)v_0/\Omega]$ are compared with simulation results for the three dominant sidebands in Fig.~\ref{Fig4}(g). 
With increasing $\omega_0$, the wavenumber difference $(k'_c - k_u)$ decreases, leading to reduced modulation depth and diminishing secondary peaks, consistent with the dispersion relation shown in Fig.~\ref{Fig2}(g). The excellent agreement between theory (solid lines) and simulation (symbols) verifies that the frequency modulation originates from nonlinear Doppler effect rather than nonlinear magnon–magnon interactions.  
This robustness is a defining hallmark of Doppler-based frequency modulation.

To explore the regime of acceleration-induced chirped Doppler shifts, we consider an incident spin wave set as a continuous $31~\mathrm{GHz}$ plane wave.
To clearly observe the chirped frequency shift induced by acceleration, multiple acceleration periods are applied according to the prescribed motion equation (see the inset of Fig.~\ref{Fig4}(h)). 
The time-dependent coordinate of the anisotropy boundary is given by
\[
\bm{x}_{\mathrm{MAB}}(t) = \left[-\frac{1}{2} \bm{a} \left(t \bmod \frac{T}{2}\right)^{2} - x_0 \right] \mathrm{sign}\!\left(t \bmod T - \frac{1}{2}\right),
\]
where $x_0 = -58~\mathrm{nm}$, the acceleration amplitude is $\bm{a} = 4|x_0|/(T/2)^2$, and the acceleration duration is $T \simeq 0.33~\mathrm{ns}$, corresponding to an effective modulation frequency of $1/T = 3~\mathrm{GHz}$. 
By extending the total observation time, a high-resolution frequency spectrum is obtained, as shown in Fig.~\ref{Fig4}(h). 
In addition to the frequency-comb characteristics, a distinct right-shifted side peak appears near the main peak, exhibiting a chirped frequency shift of $\Delta f_{\mathrm{chirp}} \simeq 1~\mathrm{GHz}$, which precisely matches the theoretical chirp bandwidth predicted by Eq.~(\ref{eq:chirpBW}). 
These results confirm that the acceleration-induced Doppler shift can be effectively utilized to probe and control the sub-boundary acceleration dynamics of spin waves.

The simulations demonstrate two complementary regimes of the nonlinear Doppler effect:  
(i)~under unbalanced torque, the oscillating boundary acts as a voltage-controlled emitter generating higher-order harmonics;  
(ii)~under balanced torque, it functions as a dispersive modulator producing frequency combs and broadband chirps.  
The excellent agreement between simulation and analytical theory across all regimes shows that these effects arise from Doppler-induced phase modulation of spin waves, establishing nonlinear boundary motion as a powerful platform for magnonic frequency synthesis and tunable frequency control.

\section{Conclusion}
In summary, we have introduced the concept of an nonlinear spin-wave Doppler effect arising from the nonlinear motion of magnetic energy boundaries in ferromagnets.
The time-dependent boundary velocity functions as an active phase modulator, producing higher-order harmonics, broadband chirps, and magnonic frequency combs—phenomena fundamentally inaccessible within conventional linear Doppler physics.  
Through micromagnetic simulations of voltage-controlled anisotropy boundaries in ferroelectric/ferromagnetic heterostructures, we have shown that both the comb spacing and overall spectral structure are governed solely by the boundary oscillation dynamics, demonstrating a direct and robust Doppler-phase coupling between boundary motion and spin-wave propagation.
The nonlinear spin-wave Doppler mechanism established here offers an energy-efficient and highly versatile route for dynamic frequency synthesis, spectral shaping, and on-chip signal transformation in magnonic circuits.  
Beyond its immediate implications for magnonic modulators and processors, this mechanism provides a new paradigm for engineering coherent magnonic spectra without relying on multi-magnon interactions or resonant mode coupling.

\newpage
\clearpage

\clearpage

\clearpage

\clearpage

\newpage
\begin{acknowledgments}
This work is supported by Shenzhen Science and Technology Program \\(JCYJ20240813113228037) and Natural Science Foundation of Top Talent of SZTU (GDRC202510). This work is also partially supported by the National Key \\Research and Development Program of China (Grant No. 2023YFB4502100, \\2020AAA0109005, 2023YFA1406600), the National Natural Science Foundation of China (Grant Nos. 62374055, 12327806, 62304083, 62074063, 61821003, 61904060, 61904051, and 61674062), the Interdisciplinary Program of Wuhan National High Magnetic Field Center (Grant No. WHMFC202119), Shenzhen Science and Technology Program Award (Grant No. JCYJ20220818103410022), Shenzhen Virtual University Park (Grant No. 2021Szvup091), Natural Science Foundation of Wuhan (Grant No. 2024040701010049).
\end{acknowledgments}

\bibliography{Ref.bib}

\end{document}